\documentclass[final,3p,twocolumn,times,sort&compress]{elsarticle}

\def\al{\alpha}
\def\be{\beta}
\def\ga{\gamma}
\def\de{\delta}
\def\ep{\epsilon}

\def\et{\eta}

\def\ka{\kappa}
\def\la{\lambda}

\def\ta{\tau}

\def\De{\Delta}

\def\Om{\Omega}

\def\cP{{\mathcal P}}
\def\cR{{\mathcal R}}

\def\fr#1#2{{{#1} \over {#2}}}
\def\half{{\textstyle{1\over 2}}}

\def\frac#1#2{{\textstyle{{#1}\over {#2}}}}

\def\lsim{\mathrel{\rlap{\lower4pt\hbox{\hskip1pt$\sim$}}
    \raise1pt\hbox{$<$}}}
\def\gsim{\mathrel{\rlap{\lower4pt\hbox{\hskip1pt$\sim$}}
    \raise1pt\hbox{$>$}}}
\def\sqr#1#2{{\vcenter{\vbox{\hrule height.#2pt
         \hbox{\vrule width.#2pt height#1pt \kern#1pt
         \vrule width.#2pt}
         \hrule height.#2pt}}}}

\def\prt{\partial}

\def\etal{{\it et al.}}

\def\pt#1{\phantom{#1}}

\def\mnab{{\mu\nu\al\be}}
\def\mn{{\mu\nu}}

\def\pr{^{\prime}}

\def\du#1{\widetilde{#1}}
\def\mx#1#2{^{#1}_{\pt{#1}#2}}

\def\kk{\ka}
\def\Q{\Om}
\def\X{X}
\def\Y{Y}
\def\tA{(T_A)}

\newcommand{\beq}{\begin{equation}}
\newcommand{\eeq}{\end{equation}}
\newcommand{\bea}{\begin{eqnarray}}
\newcommand{\eea}{\end{eqnarray}}
\newcommand{\rf}[1]{(\ref{#1})}

\begin{document}

\begin{frontmatter}

\title{Classical kinematics for Lorentz violation}

\author{V.\ Alan Kosteleck\'y$^a$ and Neil Russell$^b$}

\address{
$^a$Physics Department, Indiana University, 
Bloomington, IN 47405, USA\\
$^b$Physics Department, Northern Michigan University, 
Marquette, MI 49855, USA 
}

\address{}
\address{\rm IUHET 549, August 2010;
accepted for publication in Physics Letters B}

\begin{abstract}

Classical point-particle relativistic lagrangians are constructed 
that generate the momentum-velocity and dispersion relations
for quantum wave packets 
in Lorentz-violating effective field theory. 

\end{abstract}

\end{frontmatter}

\section{Introduction}

A promising prospect for experimental detection
of new physics at the Planck scale 
is tiny Lorentz and CPT violation
arising in an underlying unified theory such as strings
\cite{ksp}.
At experimentally attainable energy scales,
effective quantum field theory provides a useful tool
for describing observable signals 
of Lorentz and CPT violation
\cite{kp,owg}.
The comprehensive realistic effective field theory
for Lorentz and CPT violation
incorporating both the Standard Model and General Relativity 
is the Standard-Model Extension (SME) 
\cite{sme},
which has been the basis for much theoretical work 
and for numerous sensitive experimental searches
\cite{tables}.
However,
comparatively little is known about 
the corresponding classical Lorentz-violating kinematics,
a topic central to subjects such as the  
behavior of quantum wave packets,
the analysis of relativistic scattering
and the motion of macroscopic bodies.

One useful approach to the classical relativistic kinematics
of a quantum field theory
is to introduce an analogue point-particle system
with relativistic lagrangian $L$,
which leads directly to various results 
such as the equations of motion
for the classical trajectory,
the momentum-velocity connection,
and the dispersion relation.
In the Lorentz-invariant case,
the extensive literature on relativistic lagrangians
dates from Planck's 1906 work
on the free relativistic particle
\cite{mp},
Einstein's analysis of geodesics
\cite{ae},
and Frenkel's treatment of the effects of spin on trajectories
\cite{jf}.
Textbook applications of relativistic point-particle lagrangians
include the compact description of
the dispersion relations for a relativistic wave packet
and for the center-of-mass motion
of a relativistic macroscopic body,
the treatment of systems 
involving particles propagating in various spacetimes
and interacting via electromagnetic or other couplings,
and the kinematical analysis of relativistic scattering problems
\cite{ll}.

For effective field theories with Lorentz violation,
however,
available kinematical results exist primarily at leading order 
and only for simple systems.
It is known that the 3-velocity and 3-momentum
are typically misaligned
and that generic spin and momentum eigenstates
may have ill-defined velocities
\cite{sme,akrl,badc}.
The classical relativistic scattering problem
with specified inital velocities
and with interactions conserving 4-momentum
involves the explicit momentum-velocity relationship,
which is unknown for most systems with Lorentz violation.  
The kinematics of scattering in quantum field theory
requires the exact propagator 
at all orders in Lorentz violation
for external legs,
essentially because the Lorentz-invariant states 
fail to span the asymptotic Hilbert space,
and only a few processes have been analyzed 
(see, for example,
Refs.\ \cite{ck-scatt,akap,eoi,ba-compton,jlm,mfmt,fkms,hlpw}).
The study of classical Lorentz-violating trajectories 
for bodies moving under electromagnetic fields
and in post-newtonian gravity
has been restricted to leading-order terms
in Lorentz violation
\cite{traj}.
Perhaps the best understood kinematical feature 
of a Lorentz-violating quantum field theory
is the dispersion relation for a quantum wave packet. 
The form of the dispersion relation generated 
by an effective field theory is constrained
and depends on the intrinsic spin of the quantum fields
\cite{rl-disp,km-nonmin}.
In the single-fermion limit
of the renormalizable sector of the SME in Minkowski spacetime,
the exact dispersion relation has been obtained
\cite{akrl}
and techniques to study it have been developed
\cite{rl-fermion,cmm}.
In quantum electrodynamics,
the complete and exact dispersion relation for the photon,
arising from all gauge-invariant operators 
of arbitrary mass dimension,
has been constructed for uniform Lorentz violation
\cite{km-nonmin}.

In this work,
we present a method for constructing 
the classical point-particle lagrangians $L$
corresponding to a given polynomial dispersion relation 
for a particle of mass $m$ 
in the presence of uniform background fields
generically denoted by $k_x$,
where the subscript $x$ 
denotes the relevant spacetime indices.
The resulting lagrangians
describe the classical kinematics associated
with effective quantum field theories
with background fields, 
and they permit a straightforward derivation 
of exact results 
such as the momentum-velocity relation,
the dispersion relation,
and the equations of motion for the particle trajectory.
The method is directly applicable
to effective field theories with Lorentz violation,
including the SME and its various limits.
Here,
we use it to obtain some explicit lagrangians
describing the classical kinematics of the single-fermion sector 
of the minimal SME in Minkowski spacetime,
thereby resolving a number of the open issues described above. 

The single-fermion sector of the minimal SME 
contains all Lorentz-violating quadratic operators 
for a Dirac fermion of mass dimensions three and four,
which are controlled by dimension-one coefficients 
$a_\mu$, $b_\mu$, $H_\mn$
and by dimensionless coefficients
$c_\mn$, $d_\mn$, $e_\mu$, $f_\mu$, and $g_{\la\mu\nu}$,
respectively.
The corresponding exact dispersion relation 
can be obtained from the generalized Dirac equation 
for plane waves of 4-momentum $p_\mu$
by imposing that the determinant of the Dirac operator vanishes
\cite{akrl}.
It can be written in the compact form
\bea
\hskip -20pt
0 =
&&
\hskip -10pt
\frac 1 4 (V^2-S^2-A^2-P^2)^2 
\nonumber \\ &&
\hskip -10pt
+ 4 \big[P(VTA) - S(V\du T A) - VTTV + ATTA \big]
\nonumber \\ &&
\hskip -10pt
+  V^2 A^2 - (V \cdot A)^2
\nonumber \\ &&
\hskip -10pt
-  \X (V^2+S^2-A^2-P^2) - 2 \Y S P + \X^2 + \Y^2 ,
\label{smedisprel}
\eea
where
the scalar quantity is $S = -m + e \cdot p$,
the pseudoscalar is $P = f \cdot p$, 
the vector is $V_\mu = p_\mu +(cp)_\mu - a_\mu$,
the axial-vector is $A_\mu = (dp)_\mu - b_\mu$, 
and the tensor is $T_\mn = \half (g p - H)_\mn$.
The two invariants of $T_\mn$ are denoted 
$\X \equiv T_\mn T^\mn$ and
$\Y \equiv T_\mn \du T^\mn$,
with the dual defined by 
$\du T _{\mu\nu} \equiv \half \ep_{\mnab}T^{\al\be}$.
Note that for vanishing coefficients for Lorentz violation 
the dispersion relation \rf{smedisprel}
reduces to the usual form 
$(p^2 - m^2)^2 = 0$,
which is effectively quadratic.
The quadratic nature is retained 
when the only nonzero coefficients are
$a_\mu$, $c_\mn$, $e_\mu$, and $f_\mu$. 
However, the dispersion relation is generically quartic
if $b_\mu$, $d_\mn$, $g_{\la\mu\nu}$, or $H_\mn$
are nonzero.

\section{Lagrangian construction}

Consider a generic dispersion relation 
that takes the form of a polynomial equation $\cR(p_\mu;m,k_x)=0$
in the 4-momentum $p_\mu$,
with coefficients depending on the mass $m$ 
and on the background fields $k_x$.
For example,
for the single-fermion limit of the Lagrange density 
for the minimal SME,
$\cR(p)$ is the polynomial of degree four in $p_\mu$
given explicitly in Eq.\ \rf{smedisprel}.
If nonminimal quadratic Lorentz-violating operators 
of mass dimension $d$ for a single fermion are also included,
$(d-3)$ spacetime derivatives appear 
\cite{km-nonmin}.
Each row of the determinant 
of the corresponding Dirac operator in momentum space 
is then of order $(d-3)$ in $p_\mu$,
so $\cR(p)$ becomes a polynomial of degree $4(d-3)$ in $p_\mu$.

The dispersion relation $\cR(p) = 0$
expresses one on-shell condition 
on the four components of $p_\mu$.
It can be viewed as an equation 
constraining the energy $p_0$ as a function $p_0 (p_j)$
of the 3-momentum $p_j$, $j = 1,2,3$,
and it has multiple roots for $p_0$.
For background fields $k_x$ 
that are perturbative relative to the mass $m$
in any given concordant frame
\cite{akrl},
these roots include ones 
corresponding to the components of a general wave packet
decomposed into positive- and negative-energy solutions
with various spin projections.
For example,
the single-fermion limit of the minimal SME
with perturbative Lorentz violation
involves two spin projections
and hence a total of four roots $p_0$,
matching the quartic structure 
of the dispersion relation \rf{smedisprel}.
The nonminimal version of this theory
with operators up to mass dimension $d$ can have $4(d-3)$ roots,
but only four of these are perturbative.
The remaining roots correspond to high-frequency solutions
that are artifacts of effective field theory
and can be disregarded for practical purposes.
More generally,
a massive quantum field 
of intrinsic spin quantum number $j$
has $(2j+1)$ spin projections
and so the dispersion relation $\cR(p)=0$
has a total of $2(2j+1)$ perturbative roots.
Each root can be associated with
a particle or antiparticle system 
and a corresponding lagrangian $L$.
The challenge of interest here
is therefore to construct lagrangians $L$ 
for all the perturbative roots
of a given dispersion relation $\cR(p)=0$.

Using the dispersion relation $\cR(p) = 0$,
we can find three more on-shell conditions 
by taking derivatives with respect to the 3-momentum 
and requiring that 
the group 3-velocity of the quantum wave packet matches 
the 3-velocity of the classical analogue particle,
$-\prt p_0/\prt p_j = dx^j/dx^0$.
If the path of the particle is
parametrized by a path parameter $\la$,
then $dx^j/dx^0 = (dx^j/d\la)/(dx^0/d\la)$
and we can impose 
$-\prt p_0/\prt p_j = u^j/u^0$,
where $u^\mu \equiv dx^\mu/d\la$.
The three derivative conditions $\prt\cR/\prt p_j=0$
obtained from the dispersion relation
are typically nonlinear in the momenta.
For example,
the dispersion relation
for the single-fermion limit of the minimal SME
is quartic in $p_\mu$,
so the three conditions are cubic.
In the nonminimal case with operators of mass dimension $d$,
the maximum degree of the three conditions becomes $4(d-3)-1$.

To identify the point-particle theories
leading to the dispersion relation $\cR(p)= 0$
and the three conditions for $u^j$,
we seek suitable lagrangian functionals 
$L(x^\mu(\la),u^\mu(\la),\la;m,k_x)$
of the spacetime 4-position $x^\mu(\la)$
and its derivative $u^\mu(\la)$
with respect to $\la$.
We suppose that the physics of the particle system 
is independent of the path parametrization,
so $L$ has no explicit dependence on $\la$.
The action $S=\int L~d \la$
must also be reparametrization independent,
so $L$ must be homogeneous of degree one in $u^\mu$.
Invoking Euler's theorem for homogeneous functions
yields $L = -p_\mu u^\mu$,
where $p_\mu \equiv -\prt L/\prt u^\mu$
is the particle canonical 4-momentum
expressed as a function of $u^\mu$.
By construction,
this canonical 4-momentum is identified 
with the 4-momentum of the quantum wave packet.
If the background fields $k_x$ are uniform,
as in the specific examples with Lorentz violation analyzed below,
then the canonical 4-momentum $p_\mu$ is constant,
the system conserves energy and momentum,
and the lagrangians $L$ 
have no explicit dependence on $x^\mu$.

The choice of $\la$
amounts to a choice of diffeomorphism gauge 
on the one-dimensional path manifold.
On shell and for a timelike curve,
the element of path length $d\la$ 
is related to the proper-time interval $d\ta$
obeying 
$d\ta^2 \equiv \et_{\mn} dx^\mu dx^\nu$
via 
$(d\ta/d\la)^2 = u_\mu u^\mu$.
The choice of $\la$
can therefore be understood as one constraint 
on the four components $u^\mu$,
leaving three independent degrees of freedom.
Note that $\la$ can be taken as the proper time $\ta$,
in which case $u^\mu$ 
becomes the usual on-shell particle 4-velocity,
but that other choices can be more convenient
in the presence of Lorentz violation.

Collecting the above results
yields the five key equations
\beq
\cR(p) = 0, 
\quad
u^j= -u^0 \fr{\prt p_0}{\prt p_j},
\quad
L = -p_\mu u^\mu,
\label{fiveq}
\eeq
which suffice to determine $L= L(u^\mu)$.
To obtain an explicit result,
the five equations can be manipulated
to eliminate the 4-momentum components,
leaving a single equation 
that can be viewed as a polynomial for $L$,
\beq
\cP(L;u^\mu,m,k_x) \equiv \sum_{n=0}^N c_n(u^\mu;m,k_x) L^n = 0.
\label{poly}
\eeq
The degree $N$ of $\cP(L)$ 
is bounded from above by the degrees 
of the conditions \rf{fiveq}. 
For example,
we find $N\leq 12$
in the single-fermion limit of the minimal SME.
For the nonminimal case with operators of mass dimension $d$,
the degree $N$ of the polynomial $\cP(L)$
satisfies $N\leq (4d-12)(4d-13)$.

The roots of the polynomial \rf{poly}
are the candidate lagrangians $L(u^\mu;m,k_x)$
for the classical analogue systems
with positive- and negative-energy particles.
Only the roots representing perturbative deviations
from the conventional relativistic point-particle lagrangians 
$L=\pm m\sqrt{u^2}$ are of physical interest.
Other roots of $\cP(L)$ correspond
to candidate lagrangians for which the background fields are dominant
and are therefore spurious in the present context.
When the perturbative roots can be found analytically,
they generate classical lagrangians that correctly reproduce
the center-of-mass kinematics of the wave-packet components
at all orders in the background fields.

Perturbativity of the background fields
implies $p_\mu=p_\mu(u)$ can be inverted 
to yield the 4-velocity $u^\mu= u^\mu(p)$,
while the uniformity of the background fields 
ensures conservation of the canonical 4-momentum $p_\mu$
as mentioned above.
Since $L$ has no explicit dependence
on the path parameter $\la$,
these results imply that
\beq
\dot u^\mu \equiv \fr{\prt u^\mu}{\prt \la}
+ \fr{\prt u^\mu}{\prt p_\nu} \fr{d p_\nu}{d \la}
=0.
\eeq
It follows that each particle undergoes uniform motion 
in a straight line,
so Newton's first law remains unchanged 
in the presence of constant coefficients for Lorentz violation.
Effects from uniform Lorentz violation
on the behavior of a single classical point particle
that is otherwise free
are therefore unobservable {\it per se} in Minkowski spacetime,
confirming known results 
\cite{km-nonmin}.
In general,
physical effects can be detected experimentally
only by comparison of two systems 
with differing properties.
The systems may be particles of distinct flavor,
different spin projections of a single particle,
or identical particles with different momenta.
Determining the physical implications of the explicit lagrangians
obtained below therefore requires
care in establishing which two or more quantities
are being compared in a given experiment. 

We remark in passing that the above technique
for constructing lagrangians
can in principle also be used if the dispersion relation
involves nonuniform background fields,
such as those arising naturally
from curvature in Riemann spacetime
or from nonzero electric and magnetic field strengths
in quantum electrodynamics
or in the nonabelian sector. 
However,
exact dispersion relations are typically challenging 
to obtain for nonuniform fields.
Perturbative constructions such as the
Foldy-Wouthuysen method
\cite{fw}
can yield the dispersion relation to a specified order
in the background fields,
whereupon the method described here can generate
the corresponding classical point-particle lagrangians.

\section{Quadratic case}

For a massive particle and uniform background fields,
any dispersion relation quadratic in $p_\mu$
can be written in the suggestive form
\beq
(p+\kk)\Q(p+\kk) = \mu^2 ,
\label{factoredDR}
\eeq
where $\mu>0$ is a mass-like scalar,
$\kk_\mu$ is a constant 4-vector shift of the momentum,
and $\Q^\mn$ is a constant metric-like symmetric tensor.
In the limit of vanishing background fields,
$\Q^\mn \to \et^\mn$,
$\kk_\mu\to 0$,
and $\mu\to m$.
For perturbative background fields,
$\Q$ is invertible.

For this case,
calculation with the five equations \rf{fiveq}
yields a quadratic polynomial $\cP(L)$
with the root for the particle being 
\beq
L(u;\mu,\kk,\Q)
= - \mu \sqrt{u \Q^{-1} \, u} + \kk \cdot u .
\label{quadrlagr}
\eeq
The second root has the form $L(u;-\mu,\kk,\Q)$
and corresponds to the antiparticle
after reinterpretation. 
The canonical momentum for the particle is 
\beq
p_\mu \equiv - \fr{\prt L} {\prt u^\mu} =
\mu \fr {(\Q^{-1}u)_\mu} {\sqrt{u \Q^{-1}u}} - \kk_\mu .
\label{canonical}
\eeq
Notice that $p_\mu$ and $u^\mu$ generically fail to align
and that the 4-momentum $p_\mu$ 
can be nonzero when the 3-velocity vanishes,
features already noted for Lorentz violation
\cite{sme,akrl,badc}.
The dispersion relation \rf{factoredDR} can be recovered
by manipulations of Eq.\ \rf{canonical}. 
The commonly used condition $u^2 = 1$
sets the path parameter to the particle proper time,
but other choices are equally valid 
and leave the physics unaffected.
One convenient choice is 
$d \la = \sqrt{(\Q^{-1})_\mn dx^\mu dx^\nu}$,
which simplifies calculations
and matches the proper time 
in the limit of vanishing background fields.

Next,
we apply this formalism
to the SME dispersion relation \rf{smedisprel} 
restricted to nonzero coefficients 
$a_\mu$, $c_\mn$, $e_\mu$, and $f_\mu$.
For this special case,
we find $\Q = (\de + 2 c + c^T c- e e - f f)$.
The inverse $(\Q^{-1})_\mn$ can be constructed 
as an infinite series.
For $c_\mn=0$,
we find the comparatively simple lagrangian
$L(u;m,a,e,f)$ given by
\bea
\hskip -10pt
L&=& 
\hskip -2pt
-\mu 
\big\{ u^2 +
\fr 1 {\De}
\big[ (1-f^2)(e\cdot u)^2 +(1-e^2)(f\cdot u)^2 
\nonumber\\ &&
\hskip 45pt
+ 2(e\cdot f)(e\cdot u)(f \cdot u) \big] \big\}^{1/2}
- a\cdot u 
\nonumber \\ &&
+ \fr 1 \De \big[ (1-f^2)(m - e\cdot a) -(e\cdot f)(f\cdot a) \big]
e\cdot u
\nonumber \\ &&
+ \fr 1 \De \big[ (e\cdot f)(m - e\cdot a) -(1-e^2)(f\cdot a) \big]
f\cdot u ,
\nonumber \\ &&
\label{fae}
\eea
where
\bea
\mu &=& 
\fr 1 {\sqrt{\De}} \big[ (1-f^2) (m - e\cdot a)^2
\nonumber \\ &&
\hskip 20pt
- 2 (e\cdot f)(f\cdot a)(m-e\cdot a)
\nonumber \\ &&
\hskip 20pt
 + (1-e^2) (f\cdot a)^2 \big]^{1/2} ,
\eea
and where the determinant of $(\de - ee - ff)$
is $\De = (1-e^2)(1-f^2)-(e \cdot f)^2$.
If instead $c_\mn$ is the only nonzero coefficient,
then the dispersion relation reduces to
$p (\de + 2 c + c^T c)p = m^2$,
where $(c^T)_\mn=c_{\nu\mu}$.
Only the symmetric piece of the expression in parentheses 
can contribute,
and we find 
$\Q = (\de +c_S)^2 - c_A^2 +[c_S,c_A]$ ,
where
$(c_S)_\mn \equiv (c_\mn + c_{\nu\mu})/2$
and
$(c_A)_\mn \equiv (c_\mn - c_{\nu\mu})/2$.
The corresponding lagrangian is
\bea
L(u;m,c) = - m \sqrt{u \big\{
(\de +c_S)^2 - c_A^2 +[c_S,c_A]\big\}^{-1} u} .
\hskip -20pt 
\nonumber\\
\eea
The leading correction to $p_\mu$ 
therefore appears at first order if $c$ has a symmetric part
but at second order if $c$ is antisymmetric,
matching the known result at the level of field theory
\cite{sme}.

In some limits of the SME,
certain coefficients for Lorentz violation
are unphysical and can be removed by field redefinitions
\cite{sme,akrl,traj,km-nonmin,km-min,cm,mbak,ba-f,rl-b}.
For example,
the Lagrange density for a single Dirac field
with only a nonzero coefficient $a_\mu$ 
has no physical Lorentz violation
because $a_\mu$ can be eliminated
via a phase redefinition
\cite{sme}.
The classical particle then has lagrangian \rf{fae}
with $e_\mu = f_\mu = 0$,
in which $a_\mu$ appears only 
in the unphysical total-derivative term
$-a\cdot u = -d(a\cdot x)/d\la$.
When the field theory depends also on $e_\mu$ and $f_\mu$,
the required phase shift is more complicated,
but by inspection we can find it here exactly
as the coefficient of $u^\mu$ in the result \rf{fae}.
As another example,
consider the dispersion relation
$p [(\de+c)^2 - ff] p = m^2$
for symmetric $c_\mn$ and $f_\mu$.
The $f_\mu$ field can be absorbed
into a modified $c$-type coefficient $(c\pr)_\mn$
by matching the operator in brackets with $(\de+c\pr)^2$.
We find
\beq
(c\pr)\mx\mu\nu = -\de\mx\mu\nu
+\sqrt{\de\mx\mu\nu 
+ 2c\mx\mu\nu +(c^2)\mx\mu\nu - f^\mu f_\nu } ,
\eeq
to be understood as an infinite matrix series,
showing that $L(u;m,c_S,f)\equiv L(u;m,c\pr)$.
This expression reduces to the field-theoretic result
given in Ref.\ \cite{ba-f} 
for the case $c_\mn = 0$.
In the special case of a lightlike $f_\mu$ and $c_\mn = 0$,
the match $(\de+c\pr)^2 = \de - ff$ gives
$(c\pr)_\mn \equiv - \half f_\mu f_\nu$,
again in agreement with 
Ref.\ \cite{ba-f}.

\section{Quartic case}

When the coefficients
$b_\mu$, $d_\mn$, $H_\mn$, or $g_{\la\mn}$ are nonzero,
the classical properties 
of the single-fermion limit of the minimal SME
become more intractable 
and very few results are known.
Here,
we apply the above general methods 
to explore some of these cases. 

Consider first nonzero $a_\mu$ and $b_\mu$ coefficients,
for which the dispersion relation 
$\cR(p)\equiv 
[-(p-a)^2+b^2+m^2]^2 - 4[b \cdot (p-a)]^2 + 4 b^2 (p-a)^2=0$
is quartic.
Some calculations with Eq.\ \rf{fiveq}
yield an octic polynomial $\cP(L)$ of the form \rf{poly},
which factors into three pieces.
The first piece gives the two particle lagrangians 
\bea
L(u;m,a,b) = 
- m \sqrt{u^2} - a\cdot u \mp \sqrt{(b\cdot u)^2 - b^2 u^2} .
\hskip -20pt 
\nonumber\\
\label{ab:lagr2}
\eea
The second piece has the form $L(u;-m,a,b)$
and corresponds to the antiparticles 
after reinterpretation
\cite{rl-fermion,akcl},
while the third piece is spurious.
The canonical 4-momenta for $L(u;m,a,b)$ are 
\beq
p_\mu = 
\fr{m u_\mu}{\sqrt{u^2}} 
+ a_\mu 
\pm \fr{(b\cdot u) b_\mu - b^2 u_\mu}
{\sqrt{(b\cdot u)^2 - b^2 u^2}}.
\label{ab:momentum}
\eeq

The two particle lagrangians and two canonical momenta
reflect the two particle spin projections
in the quantum wave packet. 
However,
the detailed match is subtle.
Consider,
for example,
the case of timelike $b_\mu$ with particles at rest,
for which the denominator of the last term
in Eq.\ \rf{ab:momentum} vanishes. 
Choosing an observer frame in which 
$b_\mu = (b,0,0,0)$
and adopting the proper-time parametrization,
we find
$\vec p = m \vec u \mp |b| \hat u$,
revealing that
the 3-momentum and 3-velocity are collinear
but have noncoincident zeros.
We can use spatial isotropy to choose
$\vec p$ and $\vec u$ nonzero only along the 3 direction,
giving $p^3 = m u^3 \mp |b| ~\mbox{sign} (u^3)$.
For $p^3$ to be a continuous function of $u^3$,
it follows that the sign choice in Eq.\ \rf{ab:momentum}
must change when $u^3$ changes sign.

The effect of $b_\mu$ on a Dirac fermion
parallels that of minimally coupled torsion $T_{\al\be\ga}$
in a Riemann-Cartan spacetime
\cite{sme,is,krt}.
The result \rf{ab:lagr2} therefore can be adapted
to yield the analogue classical lagrangian 
for a minimally coupled Dirac field in a uniform torsion background.
For compatibility with the torsion literature,
in this paragraph we adopt the notation and conventions
of Ref.\ \cite{krt}.
The correspondence between $b_\mu$ and
the axial-vector projection 
$\tA^\mu\equiv\ep^{\al\be\ga\mu}T_{\al\be\ga}/6$
of the torsion tensor is $b_\mu=-3\tA_\mu/4$,
which yields 
\bea
L(u;T_A) = 
- m \sqrt{-u^2} \mp \frac 3 4 \sqrt{[\tA\cdot u]^2 - \tA^2 u^2} 
\hskip -20pt 
\nonumber\\
\label{torsionLagr}
\eea
as the all-orders classical lagrangian determining
the trajectory of the relativistic point particle
in a uniform axial-torsion background.

As another example with a quartic dispersion relation,
consider the case with only $H_\mn$ nonzero,
for which
$\cR(p) = 
(p^2 - m^2 + 2 \X )^2 - 8 \X p^2 - 4 p HH p + 4 \Y^2$.
Calculations with $H_\mn$ can be simplified
by noting that all nontrivial observer scalars
can be expressed in terms of $u^2$, 
the two invariants $\X \equiv H_\mn H^\mn/4$
and $\Y \equiv H_\mn \du H^\mn/4$,
and the quantity $\al\equiv uHHu$.
For example,
$u H \du H u = - \Y u^2$
and
$u H^4 u = \Y^2 u^2 - 2 \X\al$.
Also,
an observer basis can be chosen in which $(HH)\mx\mu\nu$
is diagonal with first two entries
$\sqrt{\X^2+\Y^2}-\X$
and last two entries
$-\sqrt{\X^2+\Y^2}-\X$.
This basis can be further refined 
via observer Lorentz transformations
to impose $u^1=u^2=0$,
so calculations can be performed without loss of generality
using only the two independent variables
$u^0$, $u^3$.
Nonetheless,
the general case remains refractory,
so we consider here three special instances.

First,
when $\Y=0$ some calculation shows that 
the polynomial equation $\cP(L)$ factorizes,
with roots yielding the two particle lagrangians 
\bea
\hskip -10pt
L(u;m,H;\X,\Y=0) = 
- m \sqrt{u^2} \pm \sqrt{u H H u +2 \X u^2} ,
\hskip -20pt 
\nonumber\\
\label{m_tau_lagr}
\eea
along with $L(u;-m,H;\X,\Y=0)$
corresponding to the two antiparticle solutions.
The particle canonical 4-momenta are  
\beq
p_\mu
= \fr{m u_\mu}{\sqrt{u^2}} \mp \fr{(u H H)_\mu +2 \X u_\mu}
{\sqrt{u H H u +2 \X u^2}} .
\label{H:m_sigma_mom}
\eeq
The 3-momentum and 3-velocity are typically noncollinear
and their zeros noncoincident.
When $X<0$,
the dispersion relation can be solved for $p_0$ to give
\bea
\hskip -5pt 
p_0 = 
\sqrt{\left(\sqrt{(p_2)^2+(p_3)^2} 
\pm \sqrt{-2\X}\right)^2+m^2 + (p_1)^2}
\hskip -20pt 
\nonumber\\
\eea
for the two positive sheets.
The structure is similar 
to that reported in the case of timelike $b_\mu$
\cite{akrl},
except that the sheets touch 
when the canonical momentum vanishes in the 2-3 plane
rather than in all three momentum directions.
Since the derivatives $\prt p_0/\prt p_j$
are nonexistent at zero $p_j$,
the energy-momentum space cannot be a manifold.
An interesting open question is whether
introducing an additional spin-analogue variable
would resolve this singularity.

As the second special instance of the case with nonzero $H_\mn$,
consider $\X=0$.
Some calculation reveals that $\cP(L)$ becomes quartic in $L^2$,
\bea
0 &=&
4 \Y^2 L^8
- 4 \Y^2(3m^2 u^2 + 4 \al) L^6
\nonumber\\ &&
+ \big[
      m^4(\al^2 + 12 \Y^2 u^4)
    + 16 m^2 \Y^2 \al u^2
\nonumber\\ &&
\hskip 40pt
    + 24 \Y^2 \al^2
    - 8 \Y^4 u^4
\big] L^4
\nonumber\\ &&
- 2 \big[
      m^6 (\al^2 + 2\Y^2 u^4) u^2
    + m^4 \al(\al^2 + \Y^2 u^4)
\nonumber\\ &&
\hskip 40pt
    + 2 m^2 (5 \Y^4 u^4 - \Y^2 \al^2 ) u^2
\nonumber\\ &&
\hskip 40pt
\quad
    + 8 \Y^2 \al^3
    - 8 \Y^4 \al u^4
\big] L^2
\nonumber\\ &&
+ (m^4 + 4\Y^2)(m^2 \al u^2 - \al^2 + \Y^2 u^4)^2 .
\label{sig}
\eea
All eight solutions for $L(u;m,H;\X=0,\Y)$
can be found using the standard solution for 
the roots of a quartic.
Only the perturbative roots are of interest,
corresponding to the two particle lagrangians 
and their antiparticle partners,
but their explicit form is cumbersome.
This example offers some intuition about
the complexity of the classical lagrangians
leading to the complete dispersion relation \rf{smedisprel}.

The third special case is a nonzero $H_\mn$ 
with both observer invariants $\X$ and $\Y$ vanishing 
\cite{abk}.
The dispersion relation for this case is the quartic
$(p^2-m^2)^2 = 4 p HH p$.
The corresponding polynomial $\cP(L)$ 
can be obtained from Eq.\ \rf{sig} as the limit $\Y\to 0$
while noting that $\al \neq 0$.
The two particle lagrangians $L(u;m,H;\X=0,\Y=0)$
take the form of Eq.\ \rf{m_tau_lagr} with $\X\to 0$,
as expected.

\section*{Acknowledgments}

This work was supported in part by 
the United States Department of Energy 
under grant DE-FG02-91ER40661.

\end{document}